**Experimental Demonstration of High-Performance Physical Reservoir Computing with Nonlinear Interfered Spin Wave Multi-Detection**

Wataru Namiki[1], Daiki Nishioka[1,2], Yu Yamaguchi[1,2], Takashi Tsuchiya[1*], Tohru Higuchi[2], and Kazuya Terabe[1]

[1]International Center for Materials Nanoarchitectonics (WPI-MANA), National Institute for Materials Science, 305-0044, Ibaraki, Japan
[2]Faculty of Science, Tokyo University of Science, 125-8585, Tokyo, Japan
e-mail: TSUCHIYA.Takashi@nims.go.jp

Abstract
Physical reservoir computing, which is a promising method for the implementation of highly efficient artificial intelligence devices, requires a physical system with nonlinearity, fading memory, and the ability to map in high dimensions. Although it is expected that spin wave interference can perform as highly efficient reservoir computing in some micromagnetic simulations, there has been no experimental verification to date. Herein, we demonstrate reservoir computing that utilizes multi-detected nonlinear spin wave interference in an yttrium iron garnet single crystal. The subject computing system achieved excellent performance when used for hand-written digit recognition, second-order nonlinear dynamical tasks, and nonlinear autoregressive moving average (NARMA). It is of particular note that normalized mean square errors (NMSEs) for NARMA2 and second-order nonlinear dynamical tasks were $1.81 \times 10^{-2}$ and $8.37 \times 10^{-5}$, respectively, which are the lowest figures for any experimental physical reservoir so far reported. Said high performance was achieved with higher nonlinearity and the large memory capacity of interfered spin wave multi-detection.

Reservoir computing is a promising method for implementing high-efficiency artificial intelligence that can perform at lower learning costs than is possible with conventional artificial neural networks, which is due to the system having fewer learning parameters and the ability to process time series data[1-3]. These advantages can be realized in a reservoir computing system by meeting three requirements; nonlinear transformation, short-term memory (fading memory), and the ability to map in higher dimensional space. Although recently, while various physical devices (e.g., electrical circuits, electrochemistry, magnetics, optics, and robotics) have been utilized to create physical reservoir systems[4-26], some issues remain due to the high electrical power consumption, low accuracy rates, and large volumes associated with such systems. Among related research fields, the spin wave is attracting attention due to its charge-less transport method and low electric power consumption[27-30]. Recently, there has been a report of a reservoir computing system with spin wave propagation in an active-ring resonator[14-17], although its capabilities are lacking (e.g., large volume, including resonator circuit, and low memory capacity). As a way of solving such problems, it is shown that spin wave interference in ferromagnetic materials satisfies the requirement for the three features of a physical reservoir through micromagnetic simulation[31-34]. However, there have been no experimental demonstrations that show that nonlinear interference of spin waves has been applied to a reservoir computing system.

Herein, we describe the first demonstration of a physical reservoir computing system utilizing interfered spin waves. In the subject system, an yttrium iron garnet (YIG) single crystal with multi-antennas, which are for the excitation and detection of multi-spin waves, is used as a homogeneous medium. A hand-written digit recognition task and nonlinear time series data prediction tasks were performed to evaluate the performance of the subject physical reservoir system. The maximum accuracy rate for the hand-written digit recognition task was 89.6 %, which is comparable to or higher than the score of the high performance physical reservoirs reported. Memory capacity and the solvability of nonlinear autoregressive moving average (NARMA) and second-order nonlinear dynamic tasks were dramatically improved by interfered spin wave multi-detection, which was used experimentally for the first time. Minimum errors of $8.37 \times 10^{-5}$ and $1.81\times 10^{-2}$ were achieved, which are dramatically lower than the errors of any other experimental reservoir system reported to date. One of the most noteworthy points in this paper is that our reservoir computing system can predict NARMA10 more precisely than any experimental spintronics reservoir system due to its large short-term memory capacity ($C_{\mathrm{STM}} \sim 26.98$ per 100 nodes) and high nonlinearity, which were achieved utilizing interfered spin wave multi-detection.

**The measurement configuration of interfered spin waves and the concept of reservoir computing**

Figure 1(a) shows YIG single crystal deposited coplanar antennas and its experimental configuration. A static magnetic field was applied perpendicular to a YIG surface. Two excitors and two detectors are situated so as to detect multiple signals from interfered spin waves. Microwave current injected into the excitors by an arbitrary wave form generator (AWG) induces a microwave Oersted field, which drives the precession of the magnetic moments localized near the antenna. This precession is carried away from the antennas. Then, an inverse process occurs where the dynamic magnetic dipole field, produced by the precession of the moments, induces an electromotive force in the detectors, and an oscilloscope detects the induced wave forms. Here, in the experimental configuration shown in Fig.1(b), detected spin waves are affected by two factors; (1) the interference of spin waves propagating from exciters A and B and (2) the historic effect of remnant spin precession originated from traveling spin wave. These effects give nonlinearity and fading memory properties to a reservoir computing system. Figure 1(c) shows the magnetic field dependence of a voltage induced spin wave. There are obvious variations between each signal. While the amplitude of a wave packet enhances with increasing magnetic field up to 186 mT, the amplitude of wave packets above 186 mT weaken due to damping of the spin wave during propagation, as it changes from an excitor to a detector. Figure 1(d) shows the nonlinearity of an interfered spin wave. As shown in the upper panel, a spin wave traveling between Excitor A and Detector A differs from that travelling between Excitor B and Detector A. Thus, a multi-detection technique can extract a variety of signals. As shown in the lower panel, the difference between interference and a linear combination of spin waves, shown in the upper panel after a time domain of 10 ns, shows that the interfered wave exhibits nonlinearity. This nonlinearity meets one of the necessary requirements of a physical reservoir. Figure 1(e) is a schematic illustration of a reservoir computing system with an interfered spin wave. Time-series data $u(k)$ is input to the reservoir from input layer. The input data is then transformed nonlinear one due to spin wave interference and the historic effect. Thus, a detected signal is extracted as $i$-th virtual nodes $X_i(k)$, namely neurons interacting with each other in the reservoir, to map input data to high dimensional space. A reservoir readout $y(k)$ is expressed as the product of each $X_i(k)$ and learning parameters $W_i$, as follow;

$$y(k) = \sum_{i=1}^{n} W_i X_i(k), \quad (1)$$

where $n$ and $k$ are the number of nodes and the discrete time, respectively.

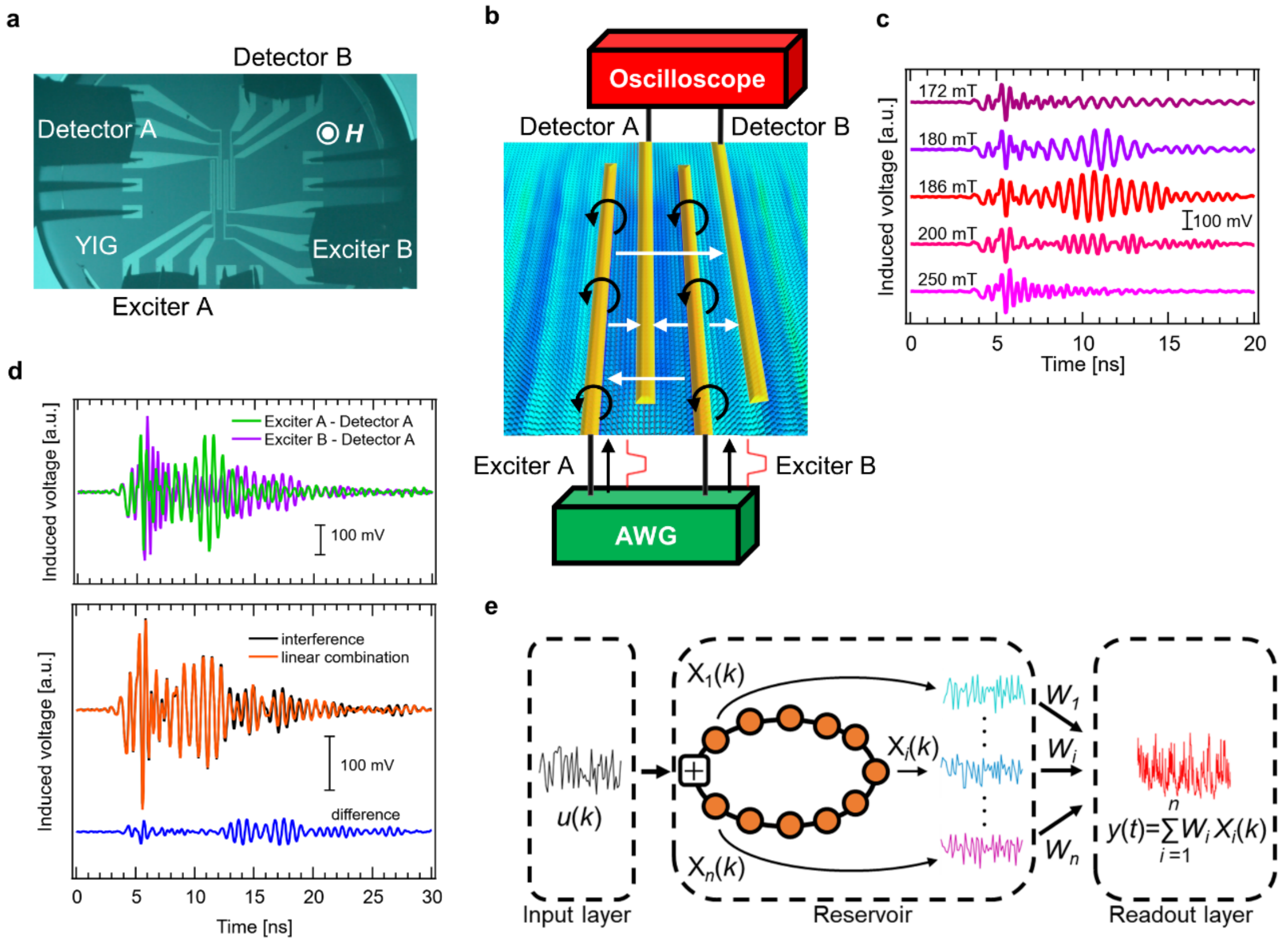

**Fig. 1| Concept of a reservoir computing system with interfered spin wave multi-detection**

**a,** YIG single crystal deposited coplanar antennas and its experimental configuration. **b,** Schematic illustration of propagation and interference of spin waves. Pulsed voltage input by AWG induces an Oersted field (solid black arrows) around the antenna. The voltage induced by the precession of magnetic moment is detected by an oscilloscope. The white arrows denote traveling spin waves excited by exciters. **c,** Induced voltage variation of spin wave packets at various magnetic fields. **d,** (Upper panel) Voltage induced by spin waves traveling between Exciter A and Detector A and between Exciter B and Detector A. (Lower panel) Compared voltage induced by an interfered spin wave and a linear combination of two spin waves. (e) General concept of a reservoir computing system with spin wave interference.

**Hand-written digit recognition task**

The first demonstration performed with our reservoir computing system was of a hand-written digit recognition task, as shown in Fig. 2(a). Hand-written digits are expressed by combinations of voltage signals with 16 different binary pulsed voltages. In the reservoir, these 16 different voltage signals are transformed to 16 different wave forms through the YIG single crystal, since excited spin waves also have various shapes. Thus, each hand-written digit is reconstructed by voltages extracted from the 16 wave forms. Figure 2(b) shows normalized induced-voltage variation at various pulsed 4-bits of data. The individual values do not overlap each other, meaning that the reservoir system gives the input data sufficient diversity. The fading memory property in the reservoir system is important for this task, since the degree of dispersion of the 16 different induced voltages is based on past input pulses in a 4-bit pulse train. Therefore, as shown in Fig. 2 (b), it is revealed that the spin wave possesses a sufficiently large fading memory property to perform the hand-written digit recognition task. Figures 2(c), (d), and (e) show the accuracy rate for hand-written digit recognition tasks with interfered spin wave, and with spin wave propagating from excitors A and B, respectively. Although the accuracy rates depend on when the reservoir state is extracted, there is no systematic dependence on time. The best accuracy rate for an interfered spin wave is 89.31 % (39 ns) at 180 mT (indicated by a black arrow). The best scoring accuracy rate for a spin wave excited at excitor A is 89.6 % (36 ns) at 172 mT (indicated by a red arrow). The best scoring accuracy rate for a spin wave excited at excitor B is 89.47 % (38 ns) at 200 mT (indicated by a light blue arrow). As shown in Fig. 2(f), the accuracy rate of 69.59 % improved as the number of training samples increased. The best score in this experiment is superior to that of any other physical reservoir computing system (e.g., memristor[5] and magnetic devices[10]) and is comparable to the excellent 91.3 % score achieved with an optical element[25], which has a larger node number of 512, and with an ionic liquid device[26].

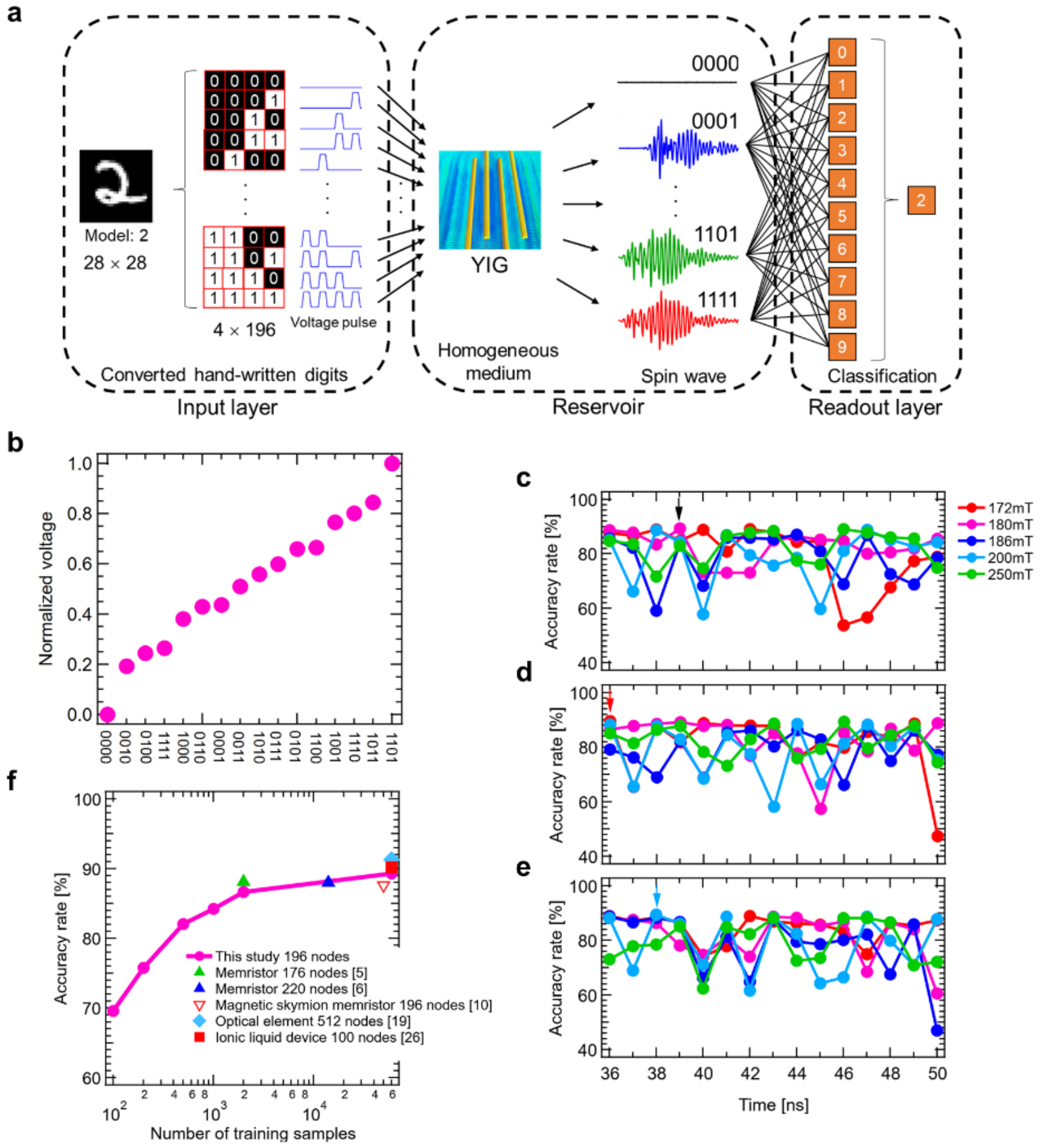

**Fig. 2| Hand-written digit recognition tasks**

**a,** General concept of a process flow diagram of a hand-written digit recognition task using a reservoir computing system with interfered spin wave. **b,** Normalized induced voltage variation at various pulsed 4-bit data. **c,d,e,** Accuracy rate variations at various times when the value of the induced voltage is extracted from an interfered spin wave **(c)**, spin wave excited from Exciter A **(d)**, and spin wave excited from Exciter B **(e)**. **f,** Number of training samples dependence of accuracy rate for hand-written digit recognition.

**Nonlinear dynamical system prediction task**

Time series data prediction tasks are widely performed so as to evaluate the nonlinear transform function of a reservoir system. The process flow for such task is shown in Fig. 3(a). In a second-order nonlinear prediction task, a random wave is input to a second-order nonlinear dynamical system. The output $d$ $(k)$ from this dynamical system at $k$ is described as follow;

$$d(k) = 0.4d(k-1) + 0.4d(k-1)d(k-2) + 0.6u^3(k) + 0.1. \quad (2)$$

$d(k)$ depends on not only the current input $u(k)$ but also on the past two states $d(k$ -1) and $d(k$ -2) at discrete times $k$ -1 and $k$ -2. Second term on the right-hand side of Eq. (2) is the cross term that makes it a second-order nonlinear system. Before being input to the reservoir system, the original random wave is processed to the pulsed signal, at intervals of 2, 5, 10, 15, and 20 ns, as pre-processing. Each of these signals is input to a reservoir computing system to which perpendicular magnetic fields of 70, 90, 97.5, 150, 169, and 176 mT are being applied. Figure 3(b) shows a comparison with the theoretical output of Eq. (2) and the predicted output reconstructed from the reservoir computing system in the training phase, which is measured under a magnetic field of 169 mT at an interval of 5 ns. The normalized mean square error (NMSE) for this task is described as follow,

$$NMSE = \sum_{k=1}^{T} \frac{\left(d(k) - y_p(k)\right)^2}{(d(k))^2}. \quad (3)$$

Here, $T$, $d(k)$, and $y_{\mathrm{p}}(k)$ are the lengths in the training phase ($T$ = 3500) or test phase ($T$ = 500), the target signal, and the predicted signal. NMSE at the training phase is $7.66 \times 10^{-5}$. A new random input is prepared for the testing phase so as to verify that the trained reservoir computing system can predict output from Eq. (2). The compared results in the testing phase are shown in Fig. 3(c). NMSE for the testing phase exhibits a similar value of $8.37 \times 10^{-5}$. Figure 3(d) shows NMSE changes at various intervals and magnetic fields. There is a tendency for the following to occur; while NMSE in stronger magnetic fields and at shorter intervals is lower, NSME in weaker magnetic fields and at longer intervals is higher under all measurement conditions. This can be seen in the comparison shown in the two upper panels (w/o interference, Detector A and Detector B) and the two lower panels (interference, Detector A and Detector B) of the figure, where NMSE is dropped overall by utilizing interfered spin waves. NMSE does not depend on the position of the detection antenna, although NMSE does change slightly in case of both interference and non-interference. However, NMSE drastically dropped to $8.37 \times 10^{-5}$ by utilizing interfered spin wave multi-detection, as is proved by the comparison shown in the two right-hand panels (Detectors A+B) and the remaining four panels (Detector A and Detector B) in the figure. This value is much lower than the values of other physical reservoir computing systems that have been reported, in which the NMSEs of a theoretical reservoir computing system with 24 spin torque oscillators[10] were ~ $1.31 \times 10^{-3}$ and an experimental reservoir computing system with 90 metal-oxide memristors[5] were ~ $3.13 \times 10^{-3}$, respectively.

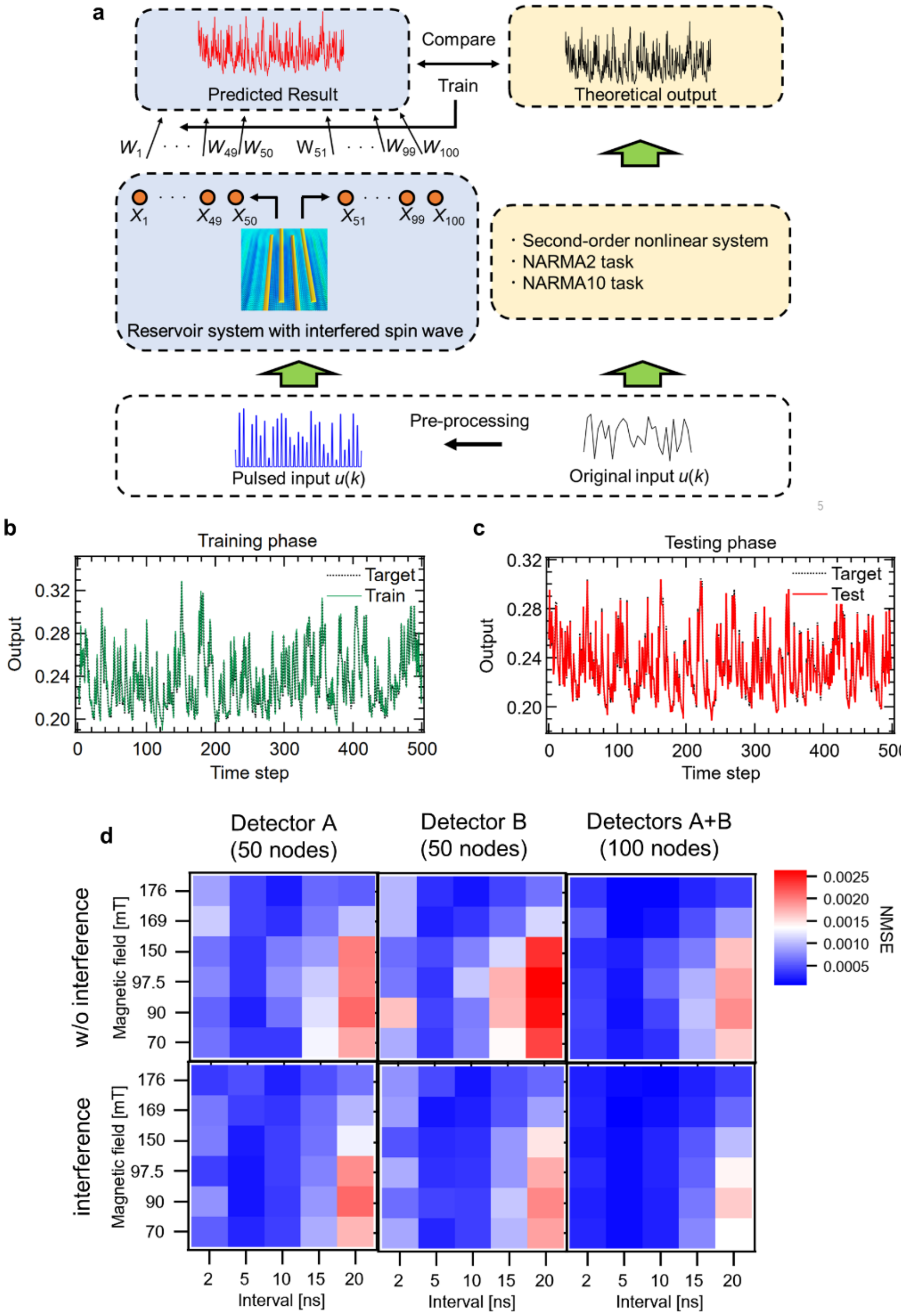

**Fig. 3| Nonlinear dynamical system prediction task**

**a,** General concept process flow diagram of a time-series prediction task using a reservoir computing system with interfered spin wave. **b,c,** Predicted results at the training phase **(b)** and testing phase **(c)** of a second order nonlinear system. The black, green, and red lines denote the target, and the prediction results at the training and testing phases, respectively. **d,** NMSE variation for prediction of a second-order nonlinear system at various magnetic fields and intervals. (Upper-left) Detector A without interference. (Upper-middle) Detector B without interference. (Upper-right) Multi-detection without interference. (Lower-left) Detector A with interference. (Lower-middle) Detector B with interference. (Lower right) Multi-detection with interference.

The nonlinear autoregressive moving average (NARMA) is a more difficult task than a second-order nonlinear dynamic task to perform, since in order to predict the output of an NARMA model, a reservoir system is required to not only perform nonlinear transform functions but also to exhibit fading memory. Here, we introduce NARMA2 and NARMA10, which need fading memory from the previous 2 and 10 steps, respectively, as defined below;

$$d(k) = 0.4d(k-1) + 0.4d(k-1)d(k-2) + 0.6u^3(k-1) + 0.1 \quad (4)$$

and

$$d(k) = 0.3d(k-1) + 0.05d(k-1)\sum_{m=1}^{10} d(k-m) + 1.5u(k-1)u(k-10) + 0.1. \quad (5)$$

Figs. 4(a) and (b) show comparisons with the theoretical outputs of Eq. (4) and predicted outputs reconstructed from the reservoir computing system in the training phase and testing phase, which are measured under a magnetic field of 169 mT and an interval of 5 ns, for NARMA2. NMSE variations at various magnetic fields and intervals are summarized in Fig. 4(c). Here, NMSE for NARMA tasks is described as follow;

$$\mathrm{NMSE_{var}} = \frac{1}{T}\sum_{k=1}^{T} \frac{\left(d(k)-y_p(k)\right)^2}{(d(k)-d_{ave.})^2}, \quad (6)$$

where $d_{\mathrm{ave.}}$ is the time average of $d(k)$. The measurement condition dependence of $\mathrm{NMSE_{var}}$ is similar to that for a second-order nonlinear dynamics task. The lowest $\mathrm{NMSE_{var}}$ for this task was $1.81 \times 10^{-2}$, which is dramatically lower than that of an experimental physical reservoir computing system previously reported[7-9]. Thus, reservoir computing systems with interfered spin wave multi-detection shows excellent computational performance. Figures 4(d) and (e) show comparisons with the theoretical outputs of Eq. (5) and predicted outputs reconstructed from the reservoir computing system in the training phase and the testing phase, which are measured under a magnetic field of 176 mT and an interval of 20 ns, for NARMA10. The $\mathrm{NMSE_{var}}$ for NARMA10 was $2.43 \times 10^{-1}$ in the testing phase under a magnetic field of 176 mT and an interval of 20 ns when utilizing interfered spin wave multi-detection. Figure 4(f) shows the NMSE variation at various intervals and magnetic fields. There is tendency, which differs from tendency for the NMSE for the NARMA10 prediction, which tendency is described as follows; while the NMSE is lower in stronger magnetic fields and longer intervals, the NSME at rest is higher. The comparison with the two upper panels (w/o interference, Detector A and Detector B) and the two lower panels (interference, Detector A and Detector B), shows that NMSE does not drop, even though interfered spin waves are utilized. NMSE does not depend on the position of the detection antenna, although NMSE slightly changes in the case of both interference and non-interference. However, NMSE dropped overall, and reached its lowest $\mathrm{NMSE_{var}}$ of $2.43 \times 10^{-1}$, when interfered spin wave multi-detection was utilized, as proved by the comparison between the two right-hand panels (Detectors A+B) and the remaining four panels (Detector A and Detector B) in

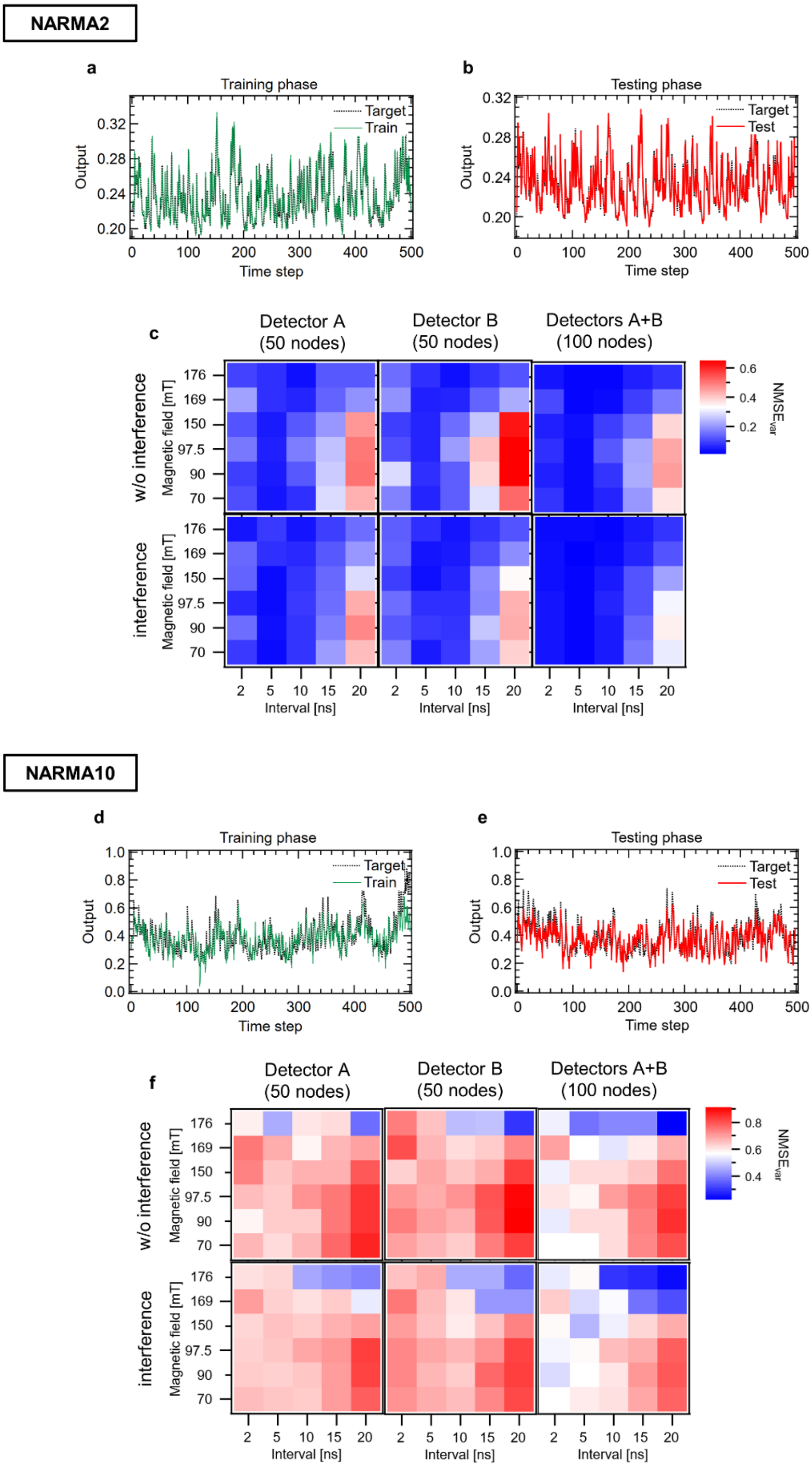

**Fig. 4| The results of NARMA2 and NARMA10 prediction tasks**

**a, b,** Prediction results at training phase **(a)** and testing phase **(b)** of the NARMA2 system. The black,

green, and red lines denote the target, and the prediction results at the training and testing phases, respectively. **c,** $NMSE_{var}$ variation for prediction of a NARMA2 system at various magnetic fields and intervals. (Upper-left) Detector A without interference. (Upper-middle) Detector B without interference. (Upper-right) Multi-detection without interference. (Lower-left) Detector A with interference. (Lower-middle) Detector B with interference. (Lower-right) Multi-detection with interference. **d, e,** Prediction results for a NARMA10 system at the training phase **(d)** and the testing phase **(e)** of a reservoir computing system with a magnetic field of 176 mT and an interval of 20ns. Thee black, green, red lines denote the target, and the prediction results at the training and testing phases, respectively. **f,** $NMSE_{var}$ variation for prediction of the NARMA10 system at various magnetic fields and intervals. (Upper-left) Detector A without interference. (Upper-middle) Detector B without interference. (Upper-right) Multi-detection without interference. (Lower-left) Detector A with interference. (Lower-middle) Detector B with interference. (Lower-right) Multi-detection with interference.

the figure. This result indicates that the large $C_{STM}$ and high ability of mapping high dimension, achieved with 100 nodes extracted from the multi-detection, is critically important in solving NAMRA10 tasks. This result confirms that a reservoir computing system with interfered spin wave multi-detection exhibits the highest performance of all experimental reservoir computing system with spintronics phenomenon ($NMSE_{var}$ of ~ $3.7 \times 10^{-1}$)[18].

**Short term memory capacity and nonlinearity of nonlinear interfered spin waves**

A short-term memory task was performed to measure the ability of our reservoir computing system for predicting time series data output from a nonlinear dynamic model. The ability for prediction decreases as the step delay increases, as shown in Fig. 5(a). This behavior accurately shows the feature of short-term memory. While a reported experimental reservoir system with spin wave propagation, which does not utilize interference and multi-detection, is able to hold the memory of 8 steps [indicated by a black arrow in Fig. 5(a)][17], the short-term memory of our reservoir system achieves longer short-term memory, above 20 steps [indicated by blue (condition I) and orange (condition II) arrows in Fig. 5(a)], by utilizing interfered spin wave multi-detection. The respective memory capacities ($C_{STM}$) under various measurement conditions is calculated from the area under the forgetting curves, as summarized in Fig. 5(b). While $C_{STM}$ drops as interval length becomes longer at the region below a magnetic field of 150 mT, $C_{STM}$ improves at the region above a magnetic field of 169 mT and an interval length of 10ns. The largest $C_{STM}$ is 26.98 per 100 nodes, which is a much larger value than the respective $C_{STM}$ below 4.85 per 20 nodes and 12.0 per 200 nodes of an experimental reservoir computing systems that utilizes spin waves with an active ring resonator and anisotropic magnetoresistance[14-18]. Thus, our reservoir system has larger $C_{STM}$ than said reservoir system with spin wave. Although the reservoir system under the condition labeled I in Fig. 5(b) can predict a NARMA10 model with the lowest $NMSE_{var}$ of $2.43 \times 10^{-1}$, the reservoir system cannot precisely predict second-order nonlinear dynamic task and NARMA2. On the other hand, the reservoir system under the condition labeled II, which has $C_{STM}$ of 13.1, can precisely predict these tasks, while that in the condition labeled I cannot, even though it has a larger $C_{STM}$ of 26.98,. This fact indicates that the system under condition I may not have enough nonlinearity to solve NARMA2.

To evaluate the nonlinearity of the system, Lyapunov exponents $\lambda$ were estimated. $\lambda$ is generally used to determine if the response of the system is orderly or disorderly (i.e., chaotic); when maximum $\lambda$ is negative (positive), the system is orderly (disorderly). Furthermore, the degree of nonlinearity of the system improves as $\lambda$ increases. As shown in Figs. 5(c) and (d), the system under the conditions labeled I and II have maximum $\lambda$ of 0.30 and 0.56, respectively. Positive maximum $\lambda$ indicates the chaotic nature of an interfered spin wave in a YIG single crystal. The condition labeled II has larger nonlinearity due to a larger maximum $\lambda$. Thus, the system under this condition has higher linearity than the condition labeled II. To eliminate the possibility of random response from the system, phase portraits of the interfered spin wave are plotted, as shown in Figs. 5(e) and (f). The orbit in the figure exhibits aperiodicity, which can also be seen in the portrait describing the chaos of the spin wave in YIG[35]. From the result in this section, there is trade-off relationship between the nonlinearity and the memory capacity of the reservoir system, which is well known as a general rule[36]. To predict second-order nonlinear tasks and NARMA2, it is important for the system to possess high nonlinearity rather than large $C_{STM}$. On the other hand, to predict NARMA10, compatibility with large $C_{STM}$ and nonlinearity is required, which is realized by utilizing interfered spin wave multi-detection. This is one of the unique advantages of the homogeneous, medium-based reservoir computing system in this study.

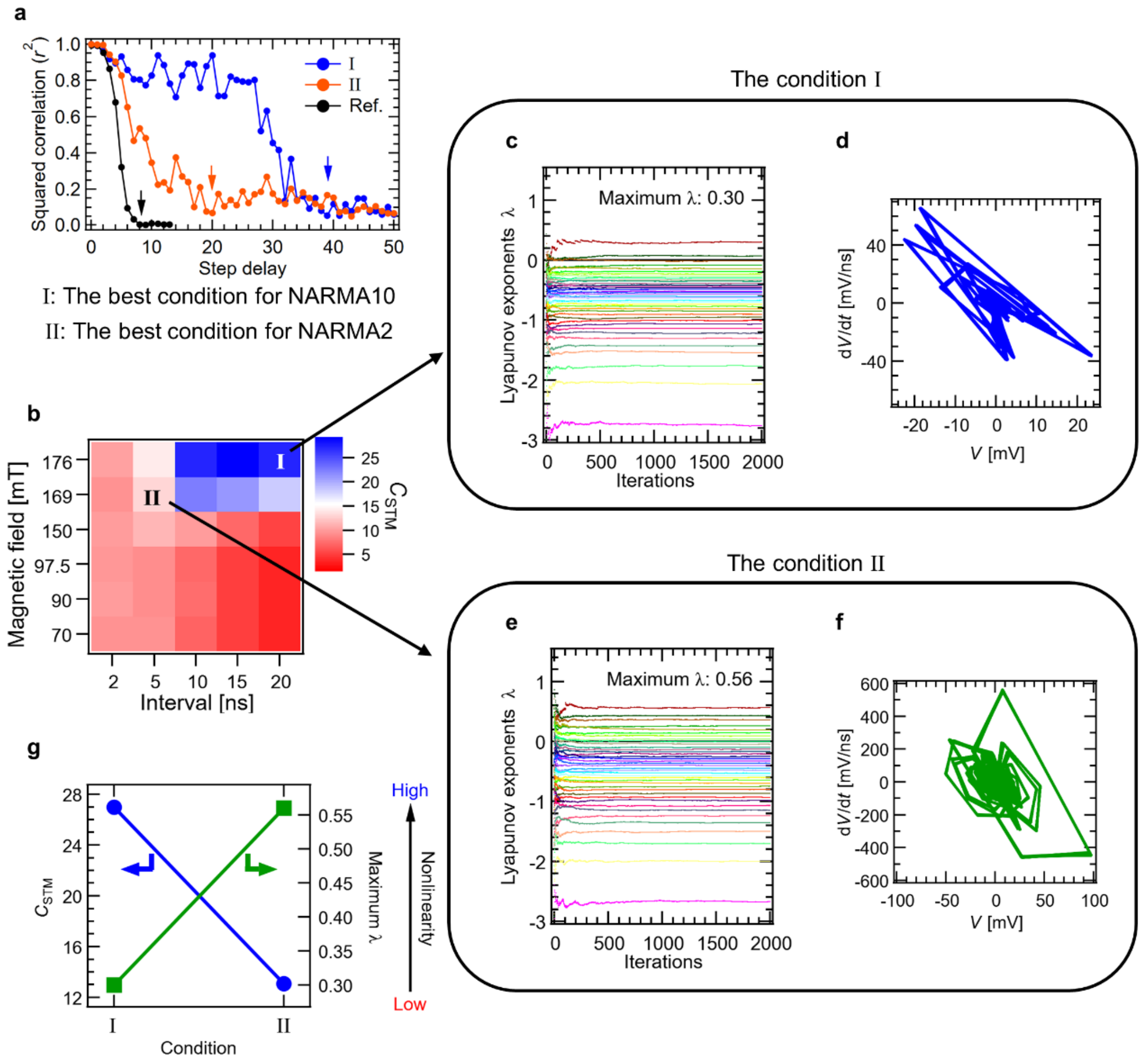

**Fig. 5| Short-term memory and nonlinearity of the reservoir system**

**a,** Forgetting curves corresponding to conditions I and II. The solid black line is the forgetting curve of a spin wave delay line active ring resonator[17]. **b,** The memory capacity variation for short term memory tasks at various magnetic fields and input intervals. I and II are the conditions that can achieve the lowest NMSE for NARMA2 and NARMA10, respectively. **c, d,** Lyapunov spectra (c) and phase portrait (d) of the conditions I. **e, f,** Lyapunov spectra (e) and phase portrait (f) of conditions II. **g,** trade-off relationship between the $C_{STM}$ and nonlinearity of an interfered spin wave system.

**Conclusions**

We achieved the first demonstration of an experimental reservoir computing system with interfered spin wave multi-detection. Our reservoir computing system achieved a recognition rate of 89.6 % for hand-written digits. The NMSEs for a second-order nonlinear dynamic task and NARMA2 were $8.37 \times 10^{-5}$ and $1.81 \times 10^{-2}$, respectively, which are dramatically lower than the NMSEs of any reservoir computing system reported to date. The subject reservoir system can also predict output from NARMA10 model, with an NMSE of $2.43 \times 10^{-1}$, which is the lowest value in an experimental spintronic reservoir computing system. While the largest $C_{STM}$ in this study was 26.98 per 100 nodes, which is larger than the $C_{STM}$ for spin wave propagation reported to date[15-17], this system has high nonlinearity, which is generally a trade-off relationship with $C_{STM}$. These high performance functions were achieved by utilizing the unique advantages of interfered spin wave multi-detection. Since this technique can be also applied to not only bulk crystal forms but also to thin film forms with extremely small volumes, the said system concept utilizes both spin wave interference and multi-detection to contribute to the implementation of integrated physical reservoir systems with practical uses.

**References**

1. Jaeger, H. & Haas, H. Harnessing nonlinearity: Predicting chaotic systems and saving energy in wireless communication. *Science* **304**, 78 (2004).
2. Maass, W., Natschläger, T. & Markram, H. Real-time computing without stable states: A new framework for neural computation based on perturbations. *Neural Comput.* **14**, 2531 (2002).
3. Verstraeten, D., Schrauwen, B., D'Haene, M. & Stroobandt, D. An experimental unification of reservoir computing methods. *Neural Networks* **20**, 391 (2007).
4. Appeltant, L., Sorano, M. C., van der Sande, G., Danckaert, J., Massar, S., Dambre, J., Schrauwen, B., Mirasso, C. R. & Fischer, I. Information processing using a single dynamical node as complex system. *Nat. Commun.* **2**, 468 (2011).
5. Du, C., Cai, F., Zidan, M. A., Ma, W., Lee, S. H. & Lu, W. D. Reservoir computing using dynamics memristors for temporal information processing. *Nat. Commun.* **8**, 2204 (2017).
6. Midya, R., Wang, A., Asapu, S., Zhang, X., Rao, M., Song, W., Zhuo, Y., Upadhyay, N., Xia, Q. & Yang, J. J. Reservoir computing using diffusive memristors. *Adv. Intel. Sys.* **1**, 1900084 (2019).
7. Kan, S., Nakajima, K., Asai, T. & Kasaya, M. A. Physical implementation of reservoir computing through electrochemical reaction. *Adv. Sci.* **9**, 2104076 (2022).
8. Kan, S., Nakajima, K., Asai, T. & Kasaya, M. A. Simple reservoir capitalizing on the nonlinear response of materials: Theory and physical implementations. *Phys. Rev. Appl.* **15**, 024030 (2021).
9. Kasaya, M. A., Takeshima, Y., Kan, S., Nakajima, K., Oya, T. & Asai, T. Performance of reservoir computing in a random network of single-walled carbon nanotubes complexed with polyoxometalate. *Neuromorph. Comput. and Eng.* **2**, 014003 (2022).
10. Jiang, W., Chen, L., Zhou, K., Li, L., Fu, Q., Du, Y. & Liu, R. H., Physical reservoir computing using magnetic skyrmion memristor and spin torque nano-oscillator. *Appl. Phys. Lett.* **115**, 192403 (2019).
11. Akashi, N., Yamaguchi, T., Tsunegi, S., Taniguchi, T., Nishida, M., Sakurai, R., Wakao, Y. & Nakajima, K. Input-driven bifurcations and information processing capacity in spintronics reservoirs. *Phys. Rev. Res.* **2**, 043303 (2020).
12. Kanao, T., Suto, H., Mizushima, K., Goto, H., Tanamoto, T. & Nagasawa, T. Reservoir computing on spin-torque oscillator array. *Phys. Rev. Appl.* **12**, 024052 (2019).
13. Tsunegi, S., Taniguchi, T., Nakajima, K., Miwa, S., Yakushiji, K., Fukushima, A., Yuasa, S. & Kubota, H. Physical reservoir computing based on spin torque oscillator with forced synchronization. *Appl. Phys. Lett.* **114**, 164101 (2019).
14. Taniguchi, T., Ogihara, A., Utsumi, Y. & Tsunegi, S. Spintronic reservoir computing without driving current or magnetic Field. *Sci. Rep.* **12**, 10627 (2022).
15. Watt, S., Kostylev, M. & Ustinov, A. B. Enhancing computational performance of a spin-wave reservoir computer with input synchronization. *J. Appl. Phys.* **129**, 044902 (2021).

16. Watt S. & Kostylev, M. Reservoir computing using a spin-wave delay-line active-ring resonator based on yttrium-iron-garnet film. *Phys. Rev. Appl.* **13**, 034057 (2020).
17. Watt, S., Kostylev, M., Ustinov, A. B. & Kalinikos, B. A. Implementing a magnonic reservoir computer model based on time-delay multiplexing. *Phys. Rev. Appl.* **15**, 064060 (2021).
18. Vidamour, I. T., Swindells, C., Venkat, G., Fry, P. W., Welbourne, A., Rowan-Robinson, R. M., Backes, D., Maccherozzi, F., Dhesi, S. S., Vasilaki, E., Allwood, D. A. & Hayward, T. J. Reservoir computing with emerging dynamics in a magnetic metameterial. *arXiv*:2206.04446 (2022).
19. Nakajima, M., Tanaka, K. & Hashimoto, T. Scalable reservoir computing on coherent linear photonic processor. *Commun. Phys.* **4**, 20 (2021).
20. Sunada, S. & Uchida, A. Photonic reservoir computing based on nonlinear wave dynamics at microscale. *Sci. Rep.* **9**, 19078 (2019).
21. Paquot, Y., Duport, F., Smerieri, A., Dambre, J., Schrauwen, B., Haelterman, M. & Massar, S. Optoelectronic reservoir computing. *Sci. Rep.* **2**, 287 (2012).
22. Duport, F., Smerieri, A., Akrout, A., Haelterman, M. & Massar, S. Fully analogue photonic reservoir computer. *Sci. Rep.* **6**, 22381 (2016).
23. Vinckier, Q., Duport, F., Smerieri, A., Vandoorne, K., Bienstman, P., Haelterman, M. & Massar, S. High-performance photonic reservoir computer based on a coherently driven passive cavity. *Optica* **2**, 438 (2015).
24. Hermans, M., Antonik, P., Haelterman, M. & Massar, S. Embodiment of learning in electro-optical signal processors. *Phys. Rev. Lett.* **117**, 128301 (2016).
25. Nakajima, K., Hauser, H., Li, T. & Pfeifer, R. Information processing via physical soft body. *Sci. Rep.* **5**, 10487 (2015).
26. Koh, S. G., Shima, H., Naitoh, Y., Akinaga, H. & Kinoshita, K. Reservoir computing with dielectric relaxation at an electrode-ionic liquid interface. *Sci. Rep.* **12**, 6958 (2022).
27. Chumak, A. V., Serga, A. A. & Hillebrands, B. Magnon transistor for all-magnon data processing. *Nat. Commun.* **5**, 4700 (2014).
28. Jamali, M., Kwon, J. H., Seo, S.-M., Lee, K.-J. & Yang, H. Spin wave nonreciprocity for logic device applications. *Sci. Rep.* **3**, 3160 (2013).
29. Klingler, S., Pirro, P., Brächer, T., Leven, B., Hillebrands, B. & Chumak, A. V. Spin-wave logic devices based on isotropic forward volume magnetostatic waves. *Appl. Phys. Lett.* **106**, 212406 (2015).
30. Goto, T., Yoshimoto, T., Iwamoto, B., Shimada, K., Ross, C. A., Sekiguchi, K., Graovsky, A. B., Nakamura, Y., Uchida, H. & Inoue, M. Three port logic gate using forward volume spin wave interference in a thin yttrium iron garnet film. *Sci. Rep.* **9**, 16472 (2019).
31. Nakane, R., Tanaka, G. & Hirose, A. Reservoir computing with spin waves excited in a garnet film. *IEEE Access* **6**, 4462 (2018).

32. Nakane, R., Hirose, A. & Tanaka, G. Spin waves propagating through a stripe magnetic domain structure and their applications to reservoir computing. *Phys. Rev. Res.* **3**, 033243 (2021).
33. Ichimura, T., Nakane, R., Tanaka, G. & Hirose, A. A numerical exploration of signal detector arrangement in a spin-wave reservoir computing device. *IEEE Access* **9**, 72637 (2021).
34. Papp, A., Csaba, G. & Porod, W. Characterization of nonlinear spin-wave interference by reservoir-computing metrics. *Appl. Phys. Lett.* **119**, 112403 (2021).
35. Gibson G. & Jeffries, C. Observation of period doubling and chaos in spin-wave instabilities in yttrium iron garnet. *Phys. Rev. A* **29**, 811 (1984).
36. Inubushi M. & Yoshimura, K. Reservoir computing beyond memory-nonlinearity trade-off. *Sci. Rep.* **7**, 10199 (2017).
37. Lecun, Y., Bottou, L., Bengio, Y. & Haffner, P. Gradient-based learning applied to document recognition. *Proceedings of the IEEE* **86**, 2278 (1998).

**Author contributions:** W.N., D.N., T.T., and K.T. conceived the idea for the study. W.N. and D.N. designed the experiments. W.N. and T.T. wrote the paper. W.N. and Y.Y. carried out the experiments. W.N. prepared the samples. W.N., D.N., T.T., and Y.Y. analyzed the data. All authors discussed the results and commented on the manuscript. K.T. directed the projects.

**Acknowledgement:**

This work was in part supported by Japan Society for the Promotion of Science (JSPS) KAKENHI Grant Number JP22H04625 (Grant-in-Aid for Scientific Research on Innovative Areas "Interface Ionics"), and JP21J21982 (Grant-in-Aid for JSPS Fellows). A part of this work was supported by the Yazaki Memorial Foundation for Science and Technology and Kurata Grants from The Hitachi Global Foundation.

## Methods

### Preparation of a device for reservoir computing system with interfered spin wave.

A one side polished-YIG single crystal with 111 orientation, which was grown by the floating zone technique, was supplied by MTI Co. (USA). Coplanar wave guides, consisting of a 10 μm wide signal line and two 20 μm wide ground lines, were fabricated by a conventional lithography technique. Ti and Au were continuously deposited by electron beam evaporator. The distance between the edges of antennas was 30 μm.

### Experimental set-up for spin wave detection.

All experiments were performed in a high frequency signal measurement system, consisting of rf probes and an electromagnet, which was made by Toei Scientific Industrial Co., Ltd.. An external magnetic field was applied perpendicular to the sample surface (i.e., the 111 direction of the YIG single crystal). The temperature of the sample was kept at room temperature (i.e., 295 ± 1 K). The excitors and detectors shown in Fig. 1(a) were connected, through rf probes, to an arbitrary waveform generator (Tektronix AWG5202) and a mixed signal digital oscilloscope (Tektronix MSO68B), respectively. Input and output signals were amplified to 30 dB and 38 dB, respectively. An average of 500 wave forms were taken to improve the signal to noise ratio. To avoid detection of excess spin waves excited by previous sequences, long 4 μs intervals were inserted between each sequence.

### Hand-written digit recognition task.

The reservoir system was trained and tested with the commonly used Mixed National Institute of Standards and Technology (MNIST) database[37]. The database contains 60,000 samples for training and 10,000 samples for testing. Each gray scale image in the dataset was comprised of 28 × 28 pixels. Original images were converted into monochrome image with the same number of pixels. Then, each the image was transformed into a 196 × 4 matrix. Each row of the matrix was encoded by a 4 bit voltage pulse, which is expressed as 16 pulse trains (i.e., 0000, 0001, 0010, … , and 1111), as the input signal for each exciting spin wave. A voltage at each time was extracted from the spin wave signal to record the device state after each pulse train. The extracted voltages were normalized, and employed a sigmoid function. Following that, 196 × 10 readout networks were obtained and used for classification. The output values were ideally equal to the target values of 1.0 for the appropriate digit and 0.0 for the remainder. In the training phase, the optimum weight coefficient vector was determined by minimizing the difference between the target and the output values. In the testing phase, the weight coefficient vector determined in the training phase was used. The accuracy rate was calculated as the ratio between the sample number classified correctly and all the sample numbers.

**Nonlinear time series data prediction task.**

The subject reservoir computing system was trained and tested with a random waveform to predict output from a NARMA model. To input to a reservoir computing system, the original random waveform $u(k)$ = [0,0.5], with time step of 5000, is transformed to pulsed waveforms with various intervals of 2 ns – 20 ns. Time steps are separated into training phases with a time step of 3500 and test phases with time step of 500 after first time step of 1000 is discarded. In the training phase, the weight parameters combining the reservoir state and the readout node are optimized so as to minimize the difference between the target waveform output $d(n)$ from the theoretical model (i.e., Eqs. (2), (4), and (5)) and the reservoir output $y(k)$. Thus, the weight coefficient $W_i$ in $y(k)$ is optimized to correspond to $d(k)$. This is described as follows;

$$y(k) = \sum_{i=1}^{n} W_i X_i(k) + b. \quad (7)$$

Here, $n$ and $b$ are the total number of nodes and a bias term, respectively. $w_i$ was optimized by ridge regression as training for the system. To compare the performance of the reservoir system with that of other systems, NMSE is calculated by using Eqs. (3) and (6).

**Estimation of short-term memory capacity.**

A delay task was performed to determine whether the system could recover past input data as current input. A random wave, with a time step of 5000, was prepared and utilized for this task. The first half of the 500 time step was discarded. The target output $d(k)$ is $u(k-\tau)$, which is a time series data delayed time step of k. The output weight coefficient $W_{out}$ was determined using the training set. The system predicts on the test set, and the square of the correlation coefficient between the ideal targets and the model predictions is determined by utilizing the relationship described as follow;

$$r^2(k) = \frac{Cov(d(k), y(k,\tau))^2}{Var(d(k)) \times Var(y(k,\tau))}, \quad (8)$$

where Cov(A, B) is the covariance between vectors A and B, and Var(A) ≅ Cov(A, A). $r^2$ takes values between 0 and 1, where the value of 1 indicates perfect replication of the targets. The short-term memory capacity $C_{STM}$ is then calculated by taking the sum of $r^2(k)$ over the range of delays. $C_{STM}$ is defined as follow;

$$C_{STM} = \sum_{k=1}^{\infty} r^2(k). \quad (9)$$

**Lyapunov spectrum.**

The Lyapunov spectrum is an index used to evaluate orbital instability, which is one of features of chaos. The Lyapunov exponent λ is defined as follow;

$$\lambda = \lim_{T \to \infty} \frac{1}{T} \sum_{t=1}^{T} \ln\left|g'(x(t))\right|. \quad (10)$$

Here, $t$ and $g'(*)$ are iteration time and differentiation of mapping function $g(*)$. If λ is positive

(negative), proximity orbits are detached (asymptotic). In calculation, the Lyapunov exponent is calculated by taking a hypersphere (ε-sphere) in m-dimensional space of minute radius ε around a point, assuming that its change after one step can be approximated linearly, and then estimating the Jacobi matrix using that variation.

The displacement vector $\boldsymbol{\mu}_i \in R^{N+L}$ with respect to $\boldsymbol{v}(k_i)$ as seen from $\boldsymbol{v}(t)$ is expressed as;

$$\boldsymbol{\mu}_i = \boldsymbol{v}(k_i) - \boldsymbol{v}(t). \quad (11)$$

Since $v(t)$ and $v(k_i)$ transit $v(t + s)$ and $v(k_i + s)$ when time $s$ elapses, the displacement vector $\boldsymbol{z}_i$ at time $t + s$ can be described as;

$$\boldsymbol{z}_i = \boldsymbol{v}(k_i + s) - \boldsymbol{v}(t + s). \quad (12)$$

Under the consumption that ε and s are small enough to be negligible, $\boldsymbol{\mu}_i$ and $\boldsymbol{z}_i$ can be linearly approximated as follow;

$$\boldsymbol{z}_i = \hat{J}(t)\boldsymbol{\mu}_i, \quad (13)$$

where $\hat{J}(t)$ is the Jacobi matrix to be estimated. Thus, the Jacobi matrix can be estimated as follows;

$$\hat{\boldsymbol{J}}(t) = \boldsymbol{z}_i\boldsymbol{\mu}_i^T(\boldsymbol{\mu}_i\boldsymbol{\mu}_i^T)^{-1}. \quad (14)$$

$\hat{J}(0)$ that is $\hat{J}(t)$ at $t = 0$ is described by QR decomposition, as follows;

$$\hat{J}(0) = \boldsymbol{Q}_1\boldsymbol{R}_1, \quad (15)$$

where **Q** and **R** are the orthogonal matrix and the upper triangular matrix. At time $t + 1$, multiply $\hat{\boldsymbol{J}}(t)$ by the orthogonal matrix from one time earlier to obtain following relation;

$$\hat{\boldsymbol{J}}(t)\boldsymbol{Q}_t = \boldsymbol{Q}_{t+1}\boldsymbol{R}_{t+1}. \quad (16)$$

Using the upper triangular matrix at each time obtained in this way, the λ is obtained as follows;

$$\lambda_i = \lim_{T\to\infty}\frac{1}{2T}\sum_{t=1}^{T}\ln\left|\boldsymbol{R}_k^{ii}\right|. \quad (17)$$

Here, $R_k^{ii}$ is the $i$-th diagonal element of the upper triangular matrix $R_k$.